\documentclass{article}
\usepackage[preprint]{spconf}
\usepackage{amsmath}
\usepackage{graphicx}
\usepackage{cite}
\usepackage{algorithmic}
\usepackage{graphicx}
\usepackage{textcomp}
\usepackage[table]{xcolor}
\usepackage{array}
\usepackage{stfloats}
\usepackage{url}
\usepackage{verbatim}
\usepackage{times}
\usepackage{epsfig}
\usepackage{amsmath}
\usepackage{amssymb}
\usepackage{amsfonts}
\usepackage{booktabs}
\usepackage{fixltx2e}
\usepackage{colortbl}
\usepackage[vlined,ruled]{algorithm2e}
\usepackage{xcolor}
\usepackage{color}
\usepackage{soul}
\usepackage{multirow}
\usepackage{float}
\usepackage{balance}
\usepackage{makecell}
\usepackage{caption}
\usepackage{subcaption}
\usepackage{lipsum}
\usepackage{rotating}
\usepackage{adjustbox}
\usepackage{pdflscape}
\usepackage{siunitx}
\usepackage{lscape}
\usepackage{pifont}
\usepackage{blindtext}
\usepackage{tabu}
\usepackage[T1]{fontenc}
\usepackage[utf8]{inputenc}
\usepackage{flexisym}
\usepackage{forest}
\usepackage{longtable}
\usepackage[multiple]{footmisc}
\usepackage{ctable}
\usepackage{epstopdf}
\usepackage{fixfoot}
\usepackage{multicol}
\usepackage{tabularray}
\usepackage{tabularx}
\usepackage[most]{tcolorbox}
\usepackage{xspace}
\usepackage{academicons}
\usepackage{soul, todonotes}
\usepackage{array}
\usepackage{fancyhdr}
\newcommand\mycomment[1]\null

\newcolumntype{P}[1]{>{\centering\arraybackslash}p{#1}}

\makeatletter
\DeclareRobustCommand\onedot{\futurelet\@let@token\@onedot}
\def\@onedot{\ifx\@let@token.\else.\null\fi\xspace}
\def\eg{\emph{e.g}\onedot} 
\def\ie{\emph{i.e}\onedot}

\def\etal{\emph{et al}\onedot}
\makeatother

\copyrightnotice{
\begin{tabular}[t]{@{}r@{}}
\copyright This work has been submitted to the IEEE for possible publication.\\Copyright may be transferred without notice, after which this version may no longer be accessible.
\end{tabular}
}

\title{HistoHDR-Net: Histogram Equalization for Single LDR to HDR Image Translation}

\name{Hrishav Bakul Barua$^{\star}$$^{\circledast}$$^{\S}$, Ganesh Krishnasamy$^{\star}$, KokSheik Wong$^{\star}$$^{\ddagger}$, Abhinav Dhall$^{\dagger}$$^{\mp}$, Kalin Stefanov$^{\dagger}$
\thanks{$^{\S}$This research is supported by the Global Excellence and Mobility Scholarship, Monash University.}
\thanks{$^{\ddagger}$This research is supported, in part, by the E-Science fund under the project: \emph{Innovative High Dynamic Range Imaging - From Information Hiding to Its Applications} (Grant No. 01-02-10-SF0327).}
}
\address{$^{\star}$School of Information Technology, Monash University Malaysia, Malaysia\\
$^{\circledast}$Robotics and Autonomous Systems Group, TCS Research, India\\
$^{\mp}$College of Science and Engineering, Flinders University, Australia\\
$^{\dagger}$Faculty of Information Technology, Monash University, Australia\\ 
\{hrishav.barua,ganesh.krishnasamy,wong.koksheik,abhinav.dhall,kalin.stefanov\}@monash.edu}

\begin{document}

\maketitle

\begin{abstract}
High Dynamic Range (HDR) imaging aims to replicate the high visual quality and clarity of real-world scenes.
Due to the high costs associated with HDR imaging, the literature offers various data-driven methods for HDR image reconstruction from Low Dynamic Range (LDR) counterparts.
A common limitation of these approaches is missing details in regions of the reconstructed HDR images, which are over- or under-exposed in the input LDR images.
To this end, we propose a simple and effective method, HistoHDR-Net, to recover the fine details (\eg, color, contrast, saturation, and brightness) of HDR images via a fusion-based approach utilizing histogram-equalized LDR images along with self-attention guidance.
Our experiments demonstrate the efficacy of the proposed approach over the state-of-art methods.
\end{abstract}


\section{Introduction}
\label{sec:introduction}
High Dynamic Range (HDR) imaging~\cite{wang2021deep, han2023high,chen2023improving} has been a topic of much focus in the vision community for the last several years.
Many applications~\cite{chen2021new,wang2021deep} require images as detailed as real-world scenes, which cannot be captured by standard Low Dynamic Range (LDR) cameras\mycomment{ (much cheaper than HDR-enabled camera systems)}.
The general camera pipeline~\cite{liu2020single} captures real-world scenes with a high dynamic range of intensity values and maps them to a low dynamic range by performing quantization, non-linear mapping, and dynamic range clipping.
Therefore, the focus in the community has been to reconstruct HDR images from their LDR counterparts using sophisticated data-driven methods that aim to approximate the reverse of the camera pipeline process.

\begin{figure}[t]
\centering
\includegraphics[width=\linewidth]{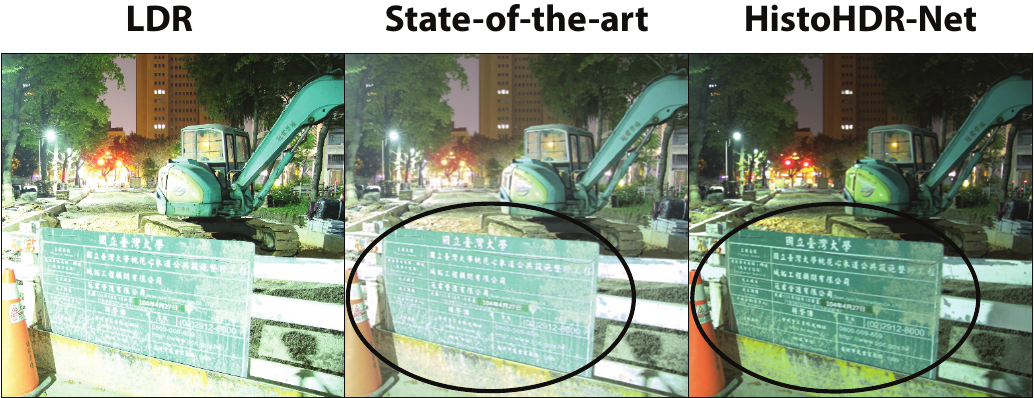}
\caption{Our method HistoHDR-Net (right) can recover the text on the notice board better than the state-of-the-art~\cite{barua2023arthdr} (middle) given an extremely over-exposed LDR image (left) as input.}
\label{fig:teaser}
\end{figure}

Most methods proposed in the literature utilize Convolutional Neural Networks and Generative Adversarial Networks~\cite{santos2020single,eilertsen2017hdr,li2019hdrnet,liu2020single,le2023single,guo2023single,zou2023rawhdr,guo2022lhdr}, where a single LDR image is used as input to reconstruct the HDR image. 
Methods based on single-exposed LDR images often produce unwanted color, contrast, saturation, hue, luminance, brightness, and radiance levels in the resulting HDR images~\cite{santos2020single,eilertsen2017hdr,li2019hdrnet,liu2020single,le2023single,guo2023single,zou2023rawhdr,guo2022lhdr}.
Some methods~\cite{niu2021hdr,cai2018learning,ren2023robust,SelfHDR} utilize multi-exposed LDR images (generally $2$ or $3$) as input but often produce many artifacts in the resultant HDR images~\cite{niu2021hdr,cai2018learning,ren2023robust,SelfHDR}. 
In contrast, ArtHDR-Net~\cite{barua2023arthdr} uses features from multi-exposed LDR images to reconstruct the final HDR images without artifacts,  but it still lacks naturality in contrast and saturation.
Recent methods employ Neural Radiance Fields~\cite{mildenhall2021nerf,huang2022hdr} drawing inspiration from physics/optics to replicate implicit color and radiance fields involved in the general camera pipeline and enable photo-realistic HDR view synthesis~\cite{mildenhall2022nerf, huang2022hdr}.
Dalal \etal~\cite{dalal2023single} used Diffusion Models and exposure loss to replicate the reverse of the camera pipeline process through a conditional diffusion architecture with a noise induction process without requiring any classifier. These advanced methods can reconstruct high-quality HDR images with increased clarity of both foreground and background.
However, further improvements are still required especially for images with highly contrasting and saturated regions.

There is limited study on the impact of histograms~\cite{yang2023lightingnet} in HDR image reconstruction.
Jang \etal~\cite{jang2020dynamic} proposed a method using histogram and color difference~\cite{coltuc2006exact} between LDR/HDR image pairs to reconstruct HDR images.
Other methods explored image fusion techniques using multi-resolution~\cite{han2023highnew} and/or multi-exposed~\cite{guo2023single} images to enhance contrast levels and remove artifacts in both the HDR and tone-mapped LDR images.
Attention-based methods have also been explored~\cite{nemoto2015visual,li2019hdrnet,zhang2019self,yan2019attention,chen2023improving}.
Li and Fang~\cite{li2019hdrnet} proposed a way to tackle the problem of color quantization using a multi-scale Convolutional Neural Network with an attention mechanism.
The method is capable of reducing the loss of information in over- and under-exposed regions.
In this work, we propose a technique that combines the capabilities of histogram equalization and data fusion to reconstruct high-quality HDR images with better saturation and contrast levels. Fig.~\ref{fig:teaser} depicts the clarity of the images generated by the proposed method compared to the recent state-of-the-art~\cite{barua2023arthdr}.
The contributions of our work, HistoHDR-Net, are:

\begin{itemize}
\item We design a simple ResNet50~\cite{shin2018cnn} based pipeline to extract features from ground truth LDR and histogram-equalized LDR images using two parallel ResNet50 blocks and perform fusion on them. We further perform self-attention on the fused feature maps before employing them for HDR image reconstruction (see Sections~\ref{sec:fusion_and_attention} and \ref{sec:atten}).
\item We design a novel loss function based on Weber's law~\cite{jiaweber}, Multi-Scale Structural Similarity Index Measure~\cite{wang2009mean,wang2004image,zhao2016loss}, and Color~\cite{sharma2005ciede2000,brainard2003color} to guide the network for better and reliable HDR image reconstruction (see Section~\ref{sec:loss}). 
\item We perform a thorough analysis of the proposed loss function, different combinations of fusion and attention mechanisms, and feature extraction backbones, to establish the contribution of each of the components (see Section~\ref{sec:results}).
\end{itemize}





\section{Method}
\label{sec:method}
\begin{figure}[t]
\centering
\includegraphics[width=\linewidth]{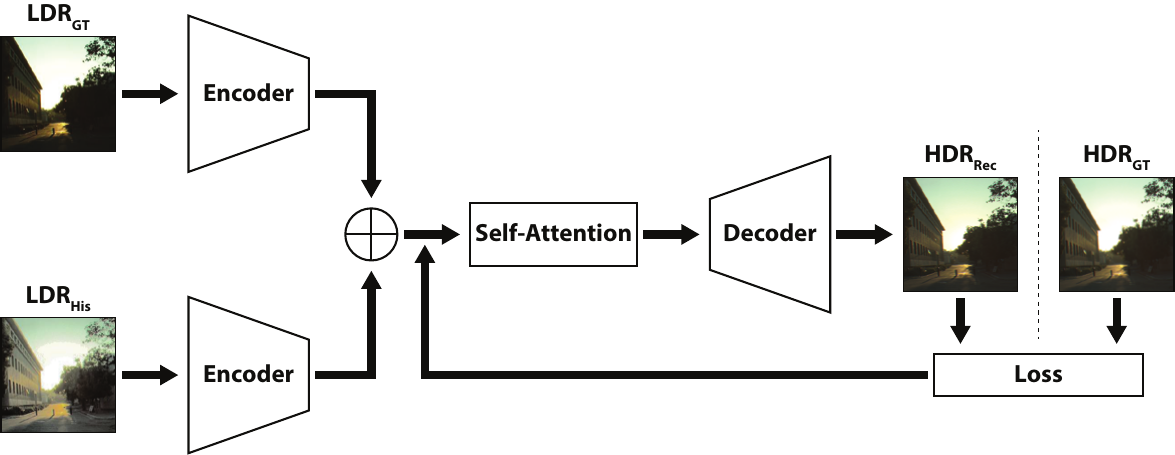}
\caption{Architecture of the proposed HistoHDR-Net method.}
\label{fig:pipeline}
\end{figure}

The proposed method employs two ResNet50~\cite{shin2018cnn} backbones for feature extraction (encoders), a feature fusion module, a self-attention block~\cite{zhang2019self} for weighing feature relevance, and a fully connected convolutions block (decoder)~\cite{khan2019fhdr,barua2023arthdr} for HDR image reconstruction.
Fig.~\ref{fig:pipeline} illustrates the proposed architecture.
We also propose a novel loss function that employs Weber's law~\cite{jiaweber} based Peak Signal-to-Noise Ratio (PSNR), Color information using the $\Delta E*$ score~\cite{sharma2005ciede2000,brainard2003color}, and Multi-Scale Structural Similarity Index Measure (MS-SSIM)~\cite{wang2009mean,wang2004image,zhao2016loss} to supervise the model in reconstructing reliable, artifact-free, and realistic HDR images with appropriate contrast and saturation levels. To the best of our knowledge Weber's law is not used as a loss in HDR reconstruction tasks so far. 

\subsection{Fusion Module}
\label{sec:fusion_and_attention}
\begin{figure}[t]
\centering
\includegraphics[width=0.8\linewidth]{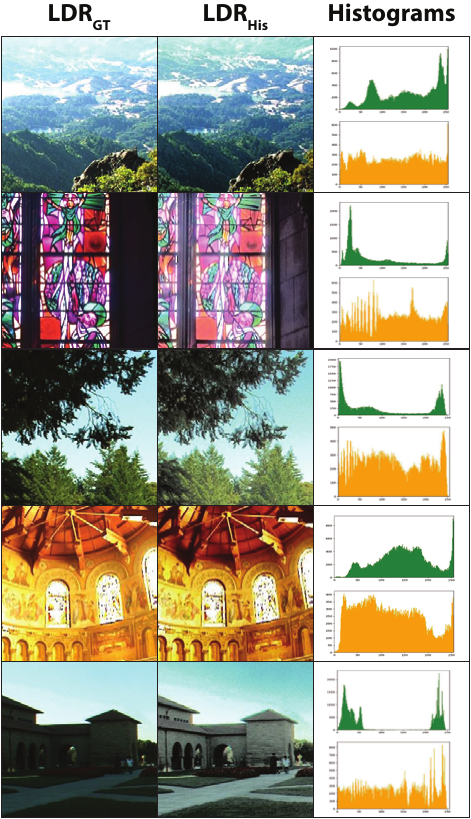}
\caption{Original and histogram-equalized LDR images. The green and orange histograms are for the original $\text{LDR}_{\text{GT}}$ and histogram-equalized $\text{LDR}_{\text{His}}$ images, respectively.}\label{fig:hist_equal}
\end{figure}

Histogram equalization is a process to improve the contrast and saturation levels.
The frequent intensity values of pixels are spread out across $256$ bins (\ie, $0-255$) which leads to intensity range equalization.
In the case of images with heavily under- and over-exposed regions, equalization may lead to revealing hidden or non-perceivable areas of the images due to extreme lighting conditions or shadows and darkness.
We use OpenCV~\cite{opencv_library} for performing histogram equalization of LDR images (see Fig.~\ref{fig:hist_equal} for examples).
We denote the original LDR images by $\text{LDR}_{\text{GT}}$ and the histogram-equalized LDR images by $\text{LDR}_{\text{His}}$.
The feature maps extracted by the two ResNet50 blocks from $\text{LDR}_{\text{GT}}$ and $\text{LDR}_{\text{His}}$ are denoted as $f^{\text{GT}}_{\text{LDR}}$ and $f^{\text{His}}_{\text{LDR}}$,  respectively. \t
Then, the output of the fusion module \ie, $f_{\text{fuse}}\text{($\cdot$)}$ can be represented as:

\begin{equation}
f_{\text{fuse}}(f^{\text{GT}}_{\text{LDR}},f^{\text{His}}_{\text{LDR}}) = f^{\text{GT}}_{\text{LDR}} \oplus f^{\text{His}}_{\text{LDR}},
\label{eq:1}
\end{equation}
where $\oplus$ represents the concatenation of feature maps. 




\subsection{Self-Attention Module}
\label{sec:atten}
The fused feature maps are passed through a simple self-attention mechanism~\cite{zhang2019self} with the intention to further improve the efficacy of the method.
This process increases the receptive field by allowing the new value of a pixel to be influenced by all other pixels in the image.
However, the main characteristic of the self-attention module is to add weights to the different features in the extracted feature maps.
Henceforth the feature importance is considered for adjusting the amount of influence it will have on the HDR image reconstruction.
We represent the output of the self-attention module as  $f_{\text{att}}\text{($\cdot$)}$.
Hence, the output of this module can be represented as $f_{\text{att}}(f_{\text{fuse}}(f^{\text{GT}}_{\text{LDR}},f^{\text{His}}_{\text{LDR}}))$.
The output of the self-attention module is added to the output of the fusion module to produce the self-attention feature map.
This information is used by the decoder block to reconstruct the HDR image from a single-exposed LDR image.

\subsection{Loss Functions}
\label{sec:loss}
All loss functions use tone-mapped versions of the reconstructed HDR ($\text{HDR}^{\text{TM}}_{\text{Rec}}$) and ground truth HDR ($\text{HDR}^{\text{TM}}_{\text{GT}}$) images.
Tone-mapping is based on the $\mu$-law~\cite{jinno2011mu} and mitigates the influence of high intensity pixel values in the HDR images\mycomment{ and prevent them from distorting the loss function calculations}.
The tone-mapping operator $T(\cdot)$ is formulated as:

\begin{equation}
T(\text{HDR}) = \frac{log(1 + \mu \text{HDR})}{log(1 + \mu)},
\label{eq:2}
\end{equation}
where $\text{HDR}^{\text{TM}}$ is the output of the tone-mapping operation $T\text{(HDR)}$ and $\mu$ specifies the amount of compression, which is set to 5000 following~\cite{khan2019fhdr}.

We designed a novel loss function consisting of five components.
The L1 ($\mathcal{L}_{L1}$) and Perceptual ($\mathcal{L}_{VGG}$) losses are commonly used in LDR to HDR reconstruction methods.
Apart from these losses, we introduce a loss based on Weber's law ($\mathcal{L}_{W}$), MS-SSIM metric ($\mathcal{L}_{SIM}$), and Color information ($\mathcal{L}_{C}$) using the $\Delta E*$ score.
The final loss is calculated as:

\begin{equation}
\mathcal{L} = \alpha\mathcal{L}_{L1} + \beta\mathcal{L}_{VGG} + \delta\mathcal{L}_{W} + \gamma\mathcal{L}_{SIM} + \lambda\mathcal{L}_{C},
\label{eq:3}
\end{equation}
where $\alpha$, $\beta$, $\delta$, $\gamma$, and $\lambda$ are scaling factors which are empirically set to $0.18$, $0.5$, $0.82$, $0.80$, and $0.82$, respectively.

To facilitate the presentation, we denote $\text{HDR}^{\text{TM}}_{\text{GT}}$ by $X$ and $\text{HDR}^{\text{TM}}_{\text{Rec}}$ by $Y$ in all loss functions.
The $\mathcal{L}_{L1}$ loss is the Mean Absolute Error between the pixels of $X$ and $Y$:

\begin{equation}
\mathcal{L}_{L1} = \frac{1}{n}\sum_{t=1}^{n} ||X_t - Y_t||,
\label{eq:4}
\end{equation}
where $n$ (in all loss functions) is the total number of images.

The $\mathcal{L}_{VGG}$ loss is also calculated between $X$ and $Y$ and uses $f_{\text{VGG}}\text{($\cdot$)}$~\cite{johson2016perceptual} and a pre-trained VGG19~\cite{simonyan2014very} model:

\begin{equation}
\mathcal{L}_{VGG} = \frac{1}{n}\sum_{t=1}^{n}f_{VGG} (X_t, Y_t).
\label{eq:5}
\end{equation}

The $\mathcal{L}_{W}$ loss uses Weber's law~\cite{jiaweber} of perceptual image meaningfulness, \ie, the background light influences the human's perception and response toward the intensity fluctuation of visual signals. The Weber's law based $\text{PSNR}_{\text{W}}$ induces the psychology of the human vision into the perceptual image quality evaluation and is defined as:

\begin{equation}
\text{PSNR}_{\text{W}} = 10\log_{10}\frac{(2^{bitDepth} - 1)^2}{\frac{1}{MN}\displaystyle \sum_{i=0}^{M-1}\sum_{j=0}^{N-1}w_{i,j}^{2}(X_{i,j} - Y_{i,j})^2},
\label{eq:6}
\end{equation}
where $w_{i,j}$ is calculated with $0.02\times(2^{bitDepth} - X_{i,j})$ using the Weber's law and $0.02$ is set using Weber's fraction~\cite{jiaweber}. 
$M$ and $N$ are image width and height in terms of pixel.
The $bitDepth$ is the depth of the pixel intensity values in the tone-mapped HDR images, in our case $bitDepth = 8$. 
Then, $\mathcal{L}_{W}$ is defined as:  

\begin{equation}
\mathcal{L}_{W} = \frac{1}{n}\sum_{t=1}^{n}\frac{1}{\text{PSNR}_{\text{W}_{t}}}.
\label{eq:7}
\end{equation}

The $\mathcal{L}_{SIM}$ loss is derived from the MS-SSIM metric for image comparison.
It operates on multiple scales using a multi-stage process of image sub-sampling.
It is a better representation of structural similarity of two images than simple SSIM.
The MS-SSIM metric is calculated as:

\begin{equation}
\text{MS-SSIM} = l_M^{\eta}\prod_{k=1}^{M}cs_k^{\tau_k},
\label{eq:8}
\end{equation}
where $l_M^{\eta}$ and $cs_k^{\tau_k}$ are defined below, $M$ is scales of operation, and $\eta$ and $\tau_k$ are set to $1$ in our implementation.
The luminance comparison between $X$ and $Y$ is calculated as:
\begin{equation}
l = \frac{2\mu_X\mu_Y + C_1}{\mu^2_X + \mu^2_Y + C_1},  
\label{eq:9}
\end{equation}
and the comparison of the contrast and the structure between $X$ and $Y$ is jointly calculated as:

\begin{equation}
cs = \frac{2\sigma_{XY} + C_2}{\sigma^2_X + \sigma^2_Y + C_2}. 
\label{eq:10}
\end{equation}
$\mu_X$ and $\mu_Y$ are the pixel sample means, $\sigma^2_X$ and $\sigma^2_Y$ are the variances, and $\sigma_{XY}$ is the covariance.
The variables $C_1 = (K_1L)^2$ and $C_2 = (K_2L)^2$ are used to gain stability in the case of small denominator division, where $K_1$ and $K_2$ are set to 0.01 and 0.03, respectively, as per the original metric definition.
$L = 2^{bitDepth} - 1$ is the representation of the dynamic range of the image pixels.
Then, the $\mathcal{L}_{SIM}$ loss is defined as:

\begin{equation}
\mathcal{L}_{SIM} = \frac{1}{n}\sum_{t=1}^{n}(1 - \text{MS-SSIM}_t).
\label{eq:11}
\end{equation}

The $\mathcal{L}_{C}$ loss is calculated on the basis of the $\Delta E*$ score and is defined as:

\begin{equation}
\mathcal{L}_{C} = \frac{1}{n}\sum_{t=1}^{n}\frac{\sqrt{\sum_{m=1}^{P}(X_{m}-Y_{m})^2}}{2^{bitDepth} - 1},
\label{eq:12}
\end{equation}
where $P$ is the total number of pixels in the images. $X_{m}$ and $Y_{m}$ are the color intensities of pixel $m$ in the corresponding image. 
$\sum_{m=1}^{P}(\cdot)$ calculates the sum of the squared differences along the color channels of the images.  


\section{Experiments}
\label{sec:experiments}
\begin{table}[t]
\scriptsize
\setlength{\tabcolsep}{1pt}
\centering
\caption{Within-dataset performance of the proposed HistoHDR-Net and state-of-the-art. The best results are in \textbf{bold} and the second best are \underline{underlined}.}
\label{Tab:quant}
\begin{tabular}{l|ccc|ccc}
\toprule[0.5mm]
\multicolumn{1}{l|}{\multirow{2}{*}{\textbf{Method}}} & \multicolumn{3}{c|}{\textbf{City Scene}~\cite{zhang2017learning}} & \multicolumn{3}{c}{\textbf{HDR-Synth \& HDR-Real}~\cite{liu2020single}}\\
\cline{2-7}\vspace{-0.2cm}\\
\multicolumn{1}{c|}{} & \textbf{PSNR}$\uparrow$ & \textbf{SSIM}$\uparrow$ & \textbf{HDR-VDP-2}$\uparrow$ & \textbf{PSNR}$\uparrow$ & \textbf{SSIM}$\uparrow$ & \textbf{HDR-VDP-2}$\uparrow$\\
\midrule[0.25mm]
FHDR~\cite{khan2019fhdr} & 32.51 & 0.90 & 67.23 & 17.11 & 0.71 & 66.72\\
SingleHDR~\cite{liu2020single} & 33.42 & 0.91 & 68.22  & 26.33 & 0.85 & 68.21\\
ArtHDR-Net~\cite{barua2023arthdr} & 35.12 & \underline{0.93} & \underline{69.31} & 33.45 & 0.88 & \underline{68.37}\\
Diffusion-based~\cite{dalal2023single} & \textbf{36.07} & \underline{0.93} & 68.51 & \textbf{33.52} & \underline{0.90} & 68.21\\
\midrule[0.25mm]
HistoHDR-Net & \underline{35.14} & \textbf{0.94} & \textbf{69.32} & \underline{33.48} & \textbf{0.91} & \textbf{69.24}\\
\bottomrule[0.5mm]
\end{tabular}
\end{table}

\begin{table}[t]
\setlength{\tabcolsep}{1pt}
\scriptsize
\centering
\caption{Cross-dataset evaluation of the proposed HistoHDR-Net and state-of-the-art on unseen dataset (HDR-Eye~\cite{nemoto2015visual}). The best results are in \textbf{bold} and the second best \underline{underlined}.}
\label{tab:zero-shot}
\begin{tabular}{l|ccc}
\toprule[0.5mm]
\textbf{Method} & \textbf{PSNR}$\uparrow$ & \textbf{SSIM}$\uparrow$ & \textbf{HDR-VDP-2}$\uparrow$\\
\midrule[0.25mm]
FHDR~\cite{khan2019fhdr} & 33.07 & 0.90   & 67.88 \\
SingleHDR~\cite{liu2020single}& 33.78 & 0.91  & 68.51 \\
ArtHDR-Net~\cite{barua2023arthdr} & 35.81 & 0.94 & \underline{69.13}\\
Diffusion-based~\cite{dalal2023single}& \textbf{36.46} & \underline{0.95} & 69.01\\
\midrule[0.25mm]
HistoHDR-Net & \underline{36.21} & \textbf{0.96} & \textbf{69.38} \\
\bottomrule[0.5mm]
\end{tabular}
\end{table}


\textbf{Implementation.} We trained the proposed method on an Ubuntu 20.04.6 LTS workstation with Intel\textregistered~Xeon\textregistered~CPU E5-2687W v3 @ 3.10GHz (20 CPU cores), 126 GB RAM (+ 2 GB swap memory), NVIDIA GeForce GTX 1080 GPU (having 8 GB memory), and 1.4 TB SSD.
The model was trained for 200 epochs.
We used a batch size of 10 and Adam optimizer~\cite{kingma2014adam}.
The learning rate was set to $1e-4$, i.e., 0.0001 at first and adjusted to decay later on.

\noindent\textbf{Datasets.} We used a mix of real and synthetic datasets to evaluate the proposed method.
The City Scene dataset~\cite{zhang2017learning} \mycomment{(\textit{ICCV 2017}) }includes 20K LDR/HDR image pairs with diverse scenes from indoor and outdoor environments, objects, and buildings.
HDR-Synth \& HDR-Real dataset~\cite{liu2020single} \mycomment{(\textit{CVPR 2020})} contains real LDR/HDR image pairs (9785 in total) and synthetic pairs (around 500).
All images have been pre-processed to $512\times512$ resolution.
These datasets were split into 80\% train and 20\% test splits and used in the within-dataset performance evaluation.
We also consider a third dataset, HDR-Eye~\cite{nemoto2015visual} having 46 real HDR images with resolution $512\times512$ for cross-dataset evaluation. For all ablation studies, we selected a subset of $5000$ random images from City Scene and HDR-Synth \& HDR-Real datasets. 

\noindent\textbf{Evaluation.} We use High Dynamic Range Visual Differences Predictor (HDR-VDP-2)~\cite{mantiuk2011hdr,narwaria2015hdr} for human level perceptual image similarity comparison.
Structural similarity of the images is measured with Structural Similarity Index Measure (SSIM)~\cite{wang2009mean,wang2004image}.
We also use the common Peak Signal-to-Noise Ratio in dB (PSNR)~\cite{gupta2011modified} metric for pixel level similarity.
We calculate the HDR-VDP-2 and SSIM scores on the ground truth and reconstructed HDR images in the linear domain.
The PSNR value is obtained on the $\mu$-law based tone-mapped ground truth and reconstructed HDR images.
We evaluate our method and compare it with four different state-of-the-art methods including, FHDR~\cite{khan2019fhdr}\mycomment{(\textit{GlobalSIP 2019})}, SingleHDR~\cite{liu2020single}\mycomment{(\textit{CVPR 2020})}, Diffusion-based~\cite{dalal2023single}\mycomment{(\textit{ICIP 2023})} and ArtHDR-Net~\cite{barua2023arthdr}\mycomment{(\textit{APSIPA 2023})}.

\section{Results and Ablations}
\label{sec:results}
\textbf{Quantitative Results.} Table~\ref{Tab:quant} summarizes the within-dataset quantitative results.
We see that the proposed method outperforms all of the selected state-of-the-art methods in terms of SSIM and HDR-VDP-2.
However, the diffusion-based model~\cite{dalal2023single} outperforms our model in terms of PSNR score on both datasets.
Table~\ref{tab:zero-shot} reports the results for cross-dataset evaluation.
The observed trend is similar to the within-dataset evaluation - our method outperforms all of the selected state-of-the-art methods in terms of SSIM and HDR-VDP-2 and second best in terms of PSNR.

\begin{figure*}[ht]
\centering
\includegraphics{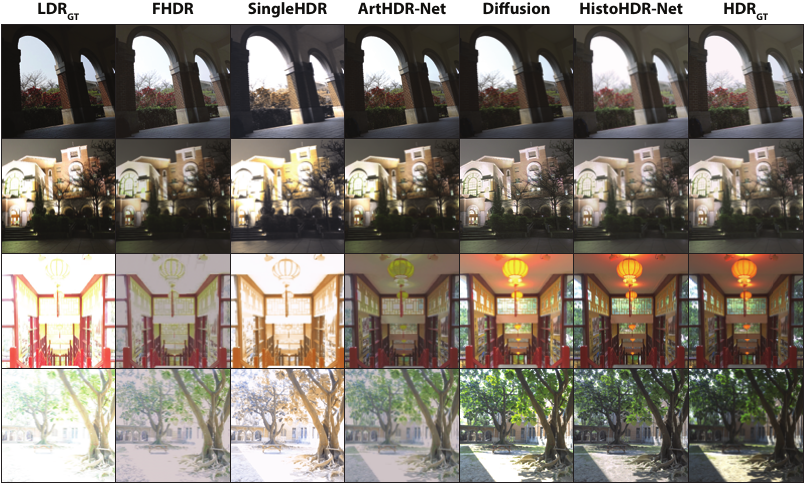}
\caption{HDR images generated by the proposed HistoHDR-Net and state-of-the-art methods.}
\label{fig:images}
\end{figure*}

\noindent \textbf{Qualitative Results.}
Fig.~\ref{fig:images} illustrates that the visual quality of our method is the closest to the ground truth HDR images in terms of color, contrast, saturation and brightness. We can also see the foreground and background objects with clarity.
We display the $\text{HDR}_{\text{GT}}$ and $\text{HDR}_{\text{Rec}}$ from all the methods using Reinhard's tone-mapping algorithm~\cite{reinhard2023photographic}.

\begin{table}[t]
\scriptsize
\centering
\caption{Architecture ablation results for different data fusion and attention approaches. `+' denotes element-wise addition and `$\oplus$' denotes concatenation. The best results are in \textbf{bold}.}
\label{tab:fusion}
\begin{tabular}{l|ccc}
\toprule[0.5mm]
\textbf{Variant} & \textbf{PSNR}$\uparrow$ & \textbf{SSIM}$\uparrow$ & \textbf{HDR-VDP-2}$\uparrow$\\
\midrule[0.25mm]
$\text{LDR}_{\text{GT}}$-only & 24.40 & 0.79 & 64.21\\
$\text{LDR}_{\text{His}}$-only & 24.11 & 0.82  & 65.02\\
\midrule[0.25mm]
$\text{LDR}_{\text{GT}}$ + $\text{LDR}_{\text{His}}$ & 26.38 & 0.85 & 65.69\\
$\text{LDR}_{\text{GT}} \oplus \text{LDR}_{\text{His}}$ & 28.11 & 0.87 & 66.12\\
\midrule[0.25mm]
$f^{\text{GT}}_{\text{LDR}}$ + $f^{\text{His}}_{\text{LDR}}$ & 30.19 & 0.88 & 67.74\\
$f^{\text{GT}}_{\text{LDR}} \oplus f^{\text{His}}_{\text{LDR}}$ & 31.43 & 0.90 & 68.01\\
\midrule[0.25mm]
$f_{\text{att}}(f^{\text{GT}}_{\text{LDR}} + f^{\text{His}}_{\text{LDR}})$ & 33.57 & 0.91 & 68.11\\
$f_{\text{att}}(f^{\text{GT}}_{\text{LDR}} \oplus f^{\text{His}}_{\text{LDR}})$ & \textbf{35.30} & \textbf{0.93} & \textbf{69.31}\\
\bottomrule[0.5mm]
\end{tabular}
\end{table}

\begin{table}[t]
\setlength{\tabcolsep}{1pt}
\scriptsize
\centering
\caption{Loss ablation results for different loss components. The best results are in \textbf{bold}.}
\label{tab:loss}
\begin{tabular}{l|ccc}
\toprule[0.5mm]
\textbf{Loss components} & \textbf{PSNR}$\uparrow$ & \textbf{SSIM}$\uparrow$ & \textbf{HDR-VDP-2}$\uparrow$\\
\midrule[0.25mm]
$\mathcal{L}_{L1}$ & 33.43 & 0.912  & 67.83\\
$\mathcal{L}_{L1}$, $\mathcal{L}_{W}$ & 34.22 & 0.917 & 68.11\\
$\mathcal{L}_{L1}$, $\mathcal{L}_{VGG}$ & 33.72 & 0.915 & 67.91\\
$\mathcal{L}_{L1}$, $\mathcal{L}_{VGG}$, $\mathcal{L}_{W}$ & 34.58 & 0.922 & 68.22\\
$\mathcal{L}_{L1}$, $\mathcal{L}_{VGG}$, $\mathcal{L}_{W}$, $\mathcal{L}_{SIM}$ & 34.98 & 0.927 & 68.85\\
$\mathcal{L}_{L1}$, $\mathcal{L}_{VGG}$, $\mathcal{L}_{W}$, $\mathcal{L}_{SIM}$, $\mathcal{L}_{C}$ & \textbf{35.30} & \textbf{0.930} & \textbf{69.31}\\
\bottomrule[0.5mm]
\end{tabular}
\end{table}

\begin{table}[t]
\scriptsize
\centering
\caption{Encoder ablation results for different feature extraction backbones. The best results are in \textbf{bold}.}
\label{tab:backbone}
\begin{tabular}{l|ccc}
\toprule[0.5mm]
\textbf{Backbone} & \textbf{PSNR}$\uparrow$ & \textbf{SSIM}$\uparrow$ & \textbf{HDR-VDP-2}$\uparrow$\\
\midrule[0.25mm]
MobileNet~\cite{howard2017mobilenets} & 32.45 & 0.895 & 67.01\\
InceptionV3~\cite{szegedy2015going} & 33.10  & 0.899 & 67.68\\
VGG19~\cite{simonyan2014very} & \textbf{35.41} & 0.927 & 69.23\\
ResNet50~\cite{shin2018cnn} & 35.30 & \textbf{0.930} & \textbf{69.31}\\
\bottomrule[0.5mm]
\end{tabular}
\end{table}

\noindent \textbf{Architecture Ablation.} We studied the impact of different fusion techniques (\ie, element-wise addition and concatenation) of the original and histogram-equalized LDR images including, early fusion (input data), late fusion (encoder feature maps), and attention mechanism.
In the case of element-wise addition, we distort the information in the original LDR with the histogram-equalized LDR images.
In contrast, concatenation allows for information sharing between the original LDR and equalized LDR images.
The results in Table~\ref{tab:fusion} demonstrate how the fusion techniques and attention mechanism improve the performance of the model compared to the baseline implementation with only LDR images or equalized LDR images as input.
The first and second rows depict the cases without fusion and attention.
We see $\text{LDR}_{\text{His}}$ is slightly better than $\text{LDR}_{\text{GT}}$.
The third and fourth rows display the early fusion results, which are slightly better than the previous case.
Histogram-equalized LDR images lead to a significant improvement when used in late fusion (fifth and sixth rows).
When self-attention is introduced to the late fusion outputs (seventh and eighth rows), we see further improvements.

\noindent \textbf{Loss Ablation.} The proposed loss function has five components out of which the first two are commonly used in image reconstruction tasks.
Table~\ref{tab:loss} shows the contribution of each of the loss components.
We see that $\mathcal{L}_{L1}$ and $\mathcal{L}_{VGG}$ achieve moderate scores and the actual improvement is achieved with the addition of $\mathcal{L}_{W}$, $\mathcal{L}_{SIM}$, and $\mathcal{L}_{C}$ (fourth, fifth, and sixth rows).
We do not use L2 loss because it is too sensitive to noise~\cite{zhao2016loss}.
We also perform a separate test on the combination of $\mathcal{L}_{L1}$ and $\mathcal{L}_{W}$ (second row).
We observe that the PSNR, SSIM and HDR-VDR-2 scores are better than the combination of $\mathcal{L}_{L1}$ and $\mathcal{L}_{VGG}$ by a margin of 1.5\%, 0.2\%, and 0.3\%, respectively.
Since, both $\mathcal{L}_{VGG}$ and $\mathcal{L}_{W}$ are perceptual losses, we can consider $\mathcal{L}_{W}$ as an alternative to $\mathcal{L}_{VGG}$ in HDR image reconstruction tasks.

\noindent \textbf{Encoder Ablation.} We also study different choices of the feature extraction block which is important for HDR image reconstruction.
We examine the most common feature extraction backbones including, MobileNet~\cite{howard2017mobilenets}, InceptionV3~\cite{szegedy2015going}, ResNet50~\cite{shin2018cnn}, and VGG19~\cite{simonyan2014very}.
The results are summarized in Table~\ref{tab:backbone}.
We see that ResNet50 attains the best results in terms of SSIM and HDR-VDP-2.
However, VGG19 reaches better performance in the case of PSNR.

\section{Conclusion}
Histogram-equalized LDR images play a significant role in improving the performance of Convolutional Neural Network based models for HDR image reconstruction.
Data fusion and self-attention mechanisms further improve the performance.
The proposed loss function that employs Weber's law, MS-SSIM metric, and $\Delta E*$ color difference further improves the reconstructed HDR images resulting in a state-of-the-art performance.
Future work includes the study of histogram matching~\cite{coltuc2006exact} that allows us to control the shape of the histogram.



\bibliographystyle{IEEEbib}
{\small
\bibliography{bibliography}}

\end{document}